\documentclass{article}
\usepackage{spconf,amsmath,epsfig}
\usepackage{amssymb}
\usepackage{multirow}
\usepackage{diagbox} 
\usepackage{tabu}
\usepackage{hyperref}
\usepackage{marvosym}
\newcommand{\nop}[1]{}

\let\OLDthebibliography\thebibliography
\renewcommand\thebibliography[1]{
  \OLDthebibliography{#1}
  \setlength{\parskip}{0pt}
  \setlength{\itemsep}{0pt plus 0.3ex}
}

\begin{document}\sloppy

\def\x{{\mathbf x}}
\def\L{{\cal L}}

\title{Factorial User Modeling with Hierarchical Graph Neural Network\\ for Enhanced Sequential Recommendation}
%

\name{Lyuxin Xue$^{\dagger}$, Deqing Yang$^{\dagger}$ \textsuperscript{\Letter} and Yanghua Xiao$^{\ddagger}$}
\address{$^{\dagger}$School of Data Science, Fudan University, Shanghai, China. \emph{\{lxxue19,yangdeqing\}@fudan.edu.cn}; \\
$^{\ddagger}$School of Computer Science, Fudan University, Shanghai, China. \emph{shawyh@fudan.edu.cn}.\\
}
\maketitle

\begin{abstract}
Most sequential recommendation (SR) systems employing graph neural networks (GNNs) only model a user's interaction sequence as a \emph{flat} graph without hierarchy, overlooking diverse factors in the user's preference. Moreover, the timespan between interacted items is not sufficiently utilized by previous models, restricting SR performance gains. To address these problems, we propose a novel SR system employing a \emph{hierarchical graph neural network} (HGNN) to model factorial user preferences. Specifically, a timespan-aware sequence graph (TSG) for the target user is first constructed with the timespan among interacted items. Next, all original nodes in TSG are softly clustered into \emph{factor nodes}, each of which represents a certain factor of the user's preference. 
At last, all factor nodes' representations are used together to predict SR results. Our extensive experiments upon two datasets justify that our HGNN-based factorial user modeling obtains better SR performance than the state-of-the-art SR models.
\end{abstract}
\begin{keywords}
Sequential recommendation, graph neural network, factorial preference, timespan
\end{keywords}
\vspace{-0.1cm}
\section{Introduction}
 \vspace{-0.2cm}
Sequential recommendation (SR) aims to leverage users' historical behaviors to predict their next interaction.
Many SR systems were built with sequential models including Markov-based models~\cite{Rendle2010Factorizing} and recurrent neural network (RNN) based models~\cite{GRUrec,Liu2018STAMP}, where a user's preference is generally represented with his/her interaction sequence. Recently, some researchers employ graph neural networks (GNNs) \cite{GCN,GGNN} to achieve SR and session-based recommendation \cite{SRGNN,GCSAN,FGNN,RetaGNN}, given that a user's historical interactions can be modeled into a \emph{sequence graph}. Accordingly, the user's dynamic preference is learned through capturing the complex transition pattern in the graph by GNNs. Despite that these models demonstrate good performance, there still exists some problems needing to be addressed. 

First, the sequence graphs in most existing GNN-based SR models are modeled as a \emph{flat graph without hierarchy}, failing to represent the diverse factors of a user's preference sufficiently. 
It has been proven that the graph embedding without hierarchy may be problematic to some downstream tasks \cite{DiffPool}. 

Second, most of previous GNN-based models \cite{SRGNN,GCSAN} built the sequence graph only with chronological order, neglecting the concrete timespan between different items in a sequence which is, however, crucial to precise SR. For example, in movie recommendation task, a user's preference on movies may vary over time. It implies that in the user's rated (or watched) movie list, the smaller timespan between two rated movies indicates potentially higher consistency and higher transition probability between them. As a result, the timespans are significant to next interaction prediction for the user. 

To address above problems, we propose a novel SR model with a \emph{hierarchical graph neural network} (HGNN), in which the representation of a user's sequence graph is learned in a \emph{hierarchical fashion} instead of flat fashion. Specifically, all (item) nodes in the graph are first clustered into several \emph{factor nodes} (super nodes) softly by the HGNN. Our empirical studies found that each factor node represents a certain factor of the user's preference, which often corresponds to a genre of items in the multimedia recommendation datasets (refer to Fig. \ref{fig:interp} in Subsec. \ref{sec:case}). Then, all factor nodes' embeddings, named as \emph{factorial preference representations}, are learned as independent as possible through adding an entropy-based regularizer in the loss function. 
At the prediction layer, these independent factorial preference representations are used together as the user's disentangled representation, to predict the interaction probability between the target user and the candidate item more precisely. 
Furthermore, the sequence graph in our model is constructed as a \emph{timespan-aware sequence graph} (TSG), where each edge is weighted by the timespan between two interacted items for SR performance gains.

In summary, our contributions in this paper include:

1. We propose HGNN to learn factorial preference representations from a user's interaction sequence graph, resulting in better user modeling with fine granularity. 

2. We design a method to incorporate timespan between different interacted items into our HGNN, based on which the SR model obtains performance gains. 

3. Our extensive experiments compared with the state-of-the-art (SOTA) SR models and ablated variants, not only justify our SR model's superior performance, but also demonstrate HGNN's advantages on differentiating the diverse factors of a user's preference and interpretability.


\vspace{-0.1cm}
\section{Methodology}
 \vspace{-0.2cm}
\subsection{Task Formalization}
 \vspace{-0.1cm}
Our task in this paper can be formalized as follows. Given a user $u$ and his/her interaction sequence $Q=\{<v_1,t_1>, <v_2,t_2>,...,<v_N,t_N>\}$ where each interaction $<v_i,t_i> (1\leq i\leq N)$ indicates that user $u$ interacted with item $v_i$ at timestamp $t_i $, the proposed model aims to predict $u$'s interacted item at the next timestamp. In general, this task is achieved through computing $u$'s probability of interacting with each candidate item $v$ given $Q$, i.e., $\hat{y}_{uv}=P(v|Q)$. 

 \vspace{-0.2cm}
\subsection{Recommendation Pipeline}
\begin{figure}[!htb]
  \centering
  \includegraphics[width=1.03\linewidth]{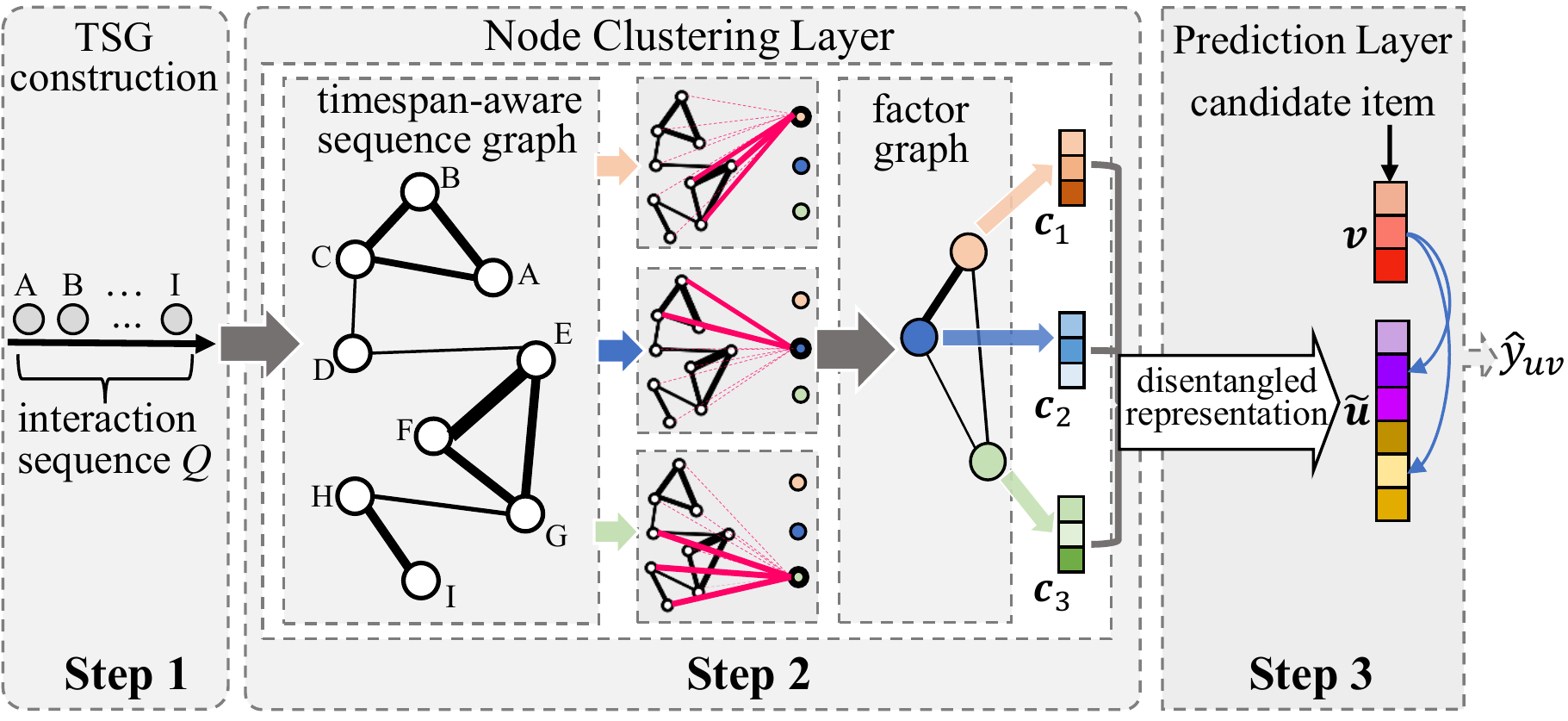}
   \vspace{-0.6cm}
  \caption{Our SR model's framework.
  }\label{fig:framework}
  \vspace{-0.2cm}
\end{figure}

Our SR's pipeline can be divided into the following three steps as depicted in Fig. \ref{fig:framework}.

\textbf{Step 1}: A timespan-aware sequence graph (TSG), denoted as $G_T$, is constructed with the input interaction sequence of the target user $u$. 

\textbf{Step 2}: 
A \emph{node clustering layer} is built to cluster the original nodes in $G_T$ into several factor nodes and outputs the factor nodes' embeddings.


\textbf{Step 3}: At last, the factorial preference representations output by Step 2 are used together as $u$'s disentangled representation, based on which $\hat{y}_{uv}$ is computed.   

 \vspace{-0.2cm}
\subsection{Timespan-aware Sequence Graph }\label{sec:TSG}
 \vspace{-0.1cm}
The $G_T$ built in our model is an undirected graph, given that the dependence pattern among the items in an interaction sequence is often \emph{bi-directional} rather than unidirectional \cite{SRGNN,MKM}. Specifically, each node in $G_T$ corresponds to an interacted item $i$ in $Q$. The timespan between two items $i, j$ in $Q$ is denoted as $\Delta t_{ij}$. We issue an edge $(i,j)$ if $\Delta t_{ij}\leq T$ where $T$ is the timespan’s upper bound. Suppose $\boldsymbol{A}$ is $G_T$'s weighted adjacency matrix, then $\boldsymbol{A}_{ij}$ is the weight of edge $(i,j)$, which is inversely proportional to the timespan and quantified as
\begin{equation}\label{eq:Aij}
\boldsymbol{A}_{ij}=
\begin{cases}
\frac{\mu}{\Delta t_{ij}}=\frac{\mu}{|t_j-t_i|}, & if \frac{\mu}{\Delta t_{ij}} \leq1;\\
1,  & otherwise.\\
\end{cases}
\notag
\end{equation}
Here $\mu$ is the timespan unit such as one day or one hour. During the information propagation and aggregation in HGNN, less information would pass between two item nodes if they are connected by an edge with big timespan or less weight. 

 \vspace{-0.2cm}
\subsection{Node Clustering Layer through HGNN}\label{sec:CL}  
 \vspace{-0.1cm}
In this step, all nodes in $G_T$ are clustered softly into several factor nodes based on the correlations among them. Specifically, an item (node) $i$ in $G_T$ is initially represented by embedding $\boldsymbol{x}_i\in \mathbb{R}^d$ which is node $i$'s feature embedding. $\boldsymbol{x}_{i}$ can be obtained through item ID projecting, if we have no item feature. Then, we encode $G_T$'s structural information into node embeddings with the similar operations in \emph{graph attention network} (GAT) \cite{GAT}, since GAT differentiates the neighbors of a node based on the attentions between them. 

The concrete operations are presented as follows. At first, $\boldsymbol{x}_i$ is linearly transformed into an embedding of $d'$ dimension as 
$
\tilde{\boldsymbol{x}_{i}} = \boldsymbol{W}^h \boldsymbol{x}_{i}
$ where $\boldsymbol{W}^h\in \mathbb{R}^{d'\times d}$ are a trainable weight matrix. 
Then, suppose $j$ is one neighbor of node $i$, we incorporate timespan $\Delta t_{ij}$ when computing the correlation between $i$ and $j$. To this end, we define
\begin{equation}\label{eq:a}
a_{ij} =\gamma\tilde{\boldsymbol{x}_{i}} \tilde{\boldsymbol{x}_{j}}^{\top} + (1-\gamma)\boldsymbol{A}_{ij}
 \vspace{-0.1cm}
\end{equation}
where $\gamma\in(0,1)$ is a control parameter. Accordingly, $a_{ij}$ consists of two factors: the feature correlation and timespan between $i$ and $j$. $\gamma$ is used to decide how much $a_{ij}$'s computation relies on either of these two factors. Then,
\begin{equation}\label{eq:alpha}
 \vspace{-0.1cm}
\begin{split}
\alpha_{ij} = \operatorname{softmax_j}\big(\operatorname{LeakyReLu}(a_{ij})\big)\\
=\frac{\exp\big(\operatorname{LeakyReLu}(a_{ij})\big)}{\sum_{k\in \mathcal{N}_i} \exp\big(\operatorname{LeakyReLu}(a_{ik})\big)}
\end{split}
 \vspace{-0.1cm}
\end{equation}
where $\mathcal{N}_{i}$ is $i$'s neighbor set. According to Eq. \ref{eq:a} and \ref{eq:alpha}, node $j$ is more correlated to node $i$ if they have similar features or smaller timespan, which conforms to SR's primary principle. 

Moreover, we adopt \emph{multi-head attention} mechanism to aggregate the information (embeddings) of $i$'s neighbors as
\begin{equation}\label{eq:z}
 \vspace{-0.1cm}
\tilde{\boldsymbol{z}_{i}}= \mathop{\Vert}\limits_{h=1}^{H}\sigma\left(\sum_{j\in \mathcal{N}_{i}}\alpha_{ij}^{h}\tilde{\boldsymbol{x}_{j}}^{h}\right),\quad
\boldsymbol{z}_{i} = \boldsymbol{W}^z \tilde{\boldsymbol{z}_{i}}
 \vspace{-0.1cm}
\end{equation}
where $H$ is the head number and $\mathop{\Vert}$ is concatenation. $\boldsymbol{W}^z\in \mathbb{R}^{d\times Hd'}$ is also a trainable weight matrix of linear transformation. Above equations show that $\alpha_{ij}$ indicates the extent of information propagation from node $j$ to node $i$. At last, $\boldsymbol{z}_{i}\in \mathbb{R}^{d}$ is $i$'s updated embedding that is refined with $G_T$'s structural information. In short, the operations from Eq. \ref{eq:a} to Eq. \ref{eq:z} are denoted as 
$
\boldsymbol{Z}=\operatorname{GAT}(\boldsymbol{A},\boldsymbol{X})
$ where $\boldsymbol{X}\in \mathbb{R}^{N \times d}$ and $\boldsymbol{Z}\in \mathbb{R}^{N \times d}$ are the initial node embedding matrix and refined node embedding matrix, respectively. 


In addition, we need to identify how to assign each node to different \emph{factor nodes}, each of which is in fact a cluster (super node). To this end, we adopt another GAT to compute the assignment distribution of all nodes. Suppose $\boldsymbol{S}\in \mathbb{R}^{N\times K}$ is the assignment matrix in which entry $\boldsymbol{S}_{ij}$ is the probability of node $i$ belonging to the $j$-th factor of $u$'s preference, and $K$ is the factor number. Specifically, we obtain $\boldsymbol{S}$ by
\begin{equation}\label{eq:S}
\boldsymbol{S} = \operatorname{softmax}\big(\operatorname{GAT}(\boldsymbol{A},\boldsymbol{Z})\boldsymbol{W}^s\big)
 \vspace{-0.1cm}
\end{equation}
where $\boldsymbol{W}^s\in \mathbb{R}^{d\times K}$ is a trainable weight matrix. Note that this clustering is indeed a soft clustering, since each row in $\boldsymbol{S}$ is a probability distribution vector instead of a one-hot vector. But the entropy-based loss of our model introduced in Eq. \ref{eq:Ldis} ensures that each node belongs to only one factor node as far as possible, resulting in distinct preference factors and better interpretability. With $\boldsymbol{S}$, we can obtain the embeddings of all $K$ factor nodes to constitute a factor embedding matrix $\boldsymbol{C}\in\mathbb{R}^{K\times d}$, which is in fact the output of the HGNN in this clustering layer, denoted as
$
\boldsymbol{C}=\operatorname{HGNN}(\boldsymbol{A},\boldsymbol{X}).
$
 Specifically, the $j$-th $(1\leq j\leq K)$ row in $\boldsymbol{C}$ is the $j$-th factor node's representation and computed as
\begin{equation}\label{eq:c}
\boldsymbol{c}_{j} = \sum\limits_{i=1}^{N}\boldsymbol{S}_{ij}\boldsymbol{z}_{i}.
\end{equation}

According to our model's design principle, each factor node in the clustered $G_T$ is as distinct (independent) as possible to each other. Thus, we can directly use all factor nodes' embeddings, i.e., factorial preference representations, to constitute $u$'s disentangled representation as
\begin{equation}\label{eq:uc}
\boldsymbol{u}=\mathop{\Vert}\limits_{j=1}^{K}\boldsymbol{c}_j=[\boldsymbol{c}_{1},\boldsymbol{c}_{2},...,\boldsymbol{c}_{K}].
 \vspace{-0.1cm}
\end{equation}

 \vspace{-0.2cm}
\subsection{Model Prediction}
 \vspace{-0.1cm}
In this step, we compute $\hat{y}_{uv}$ with $u$'s disentangled representation and the candidate item $v$'s embedding $\boldsymbol{v}\in\mathbb{R}^d$ together with relevant timespan. Given that different factors represent $u$'s preference to different extents, we first refine $u$'s representation in Eq. \ref{eq:uc} into an attentive disentangled representation as 
\begin{equation}\label{eq:uc1}
    \tilde{\boldsymbol{u}}=[\tilde{\boldsymbol{c}}_{1},\tilde{\boldsymbol{c}}_{2},...,\tilde{\boldsymbol{c}}_{K}]= [\beta_{1}\boldsymbol{c}_{1},\beta_{2}\boldsymbol{c}_{2},...,\beta_{K}\boldsymbol{c}_{K}]
\end{equation}
where weight $\beta_{j} (1\leq j\leq K)$ is computed based on the embedding and timestamp correlations between factor node $j$ and candidate item $v$. The timestamp of factor node $j$ is identified as 
\begin{equation}\label{eq:t_j}
    t_{j}=\frac{\sum\limits_{i=1}^N\boldsymbol{S}_{ij}t_i}{S}, \quad S=\sum\limits_{i=1}^N\boldsymbol{S}_{ij}
\end{equation}
where $i$ is an item node.
Then, 
\begin{equation}\label{eq:beta}
    \beta_{j}=\operatorname{softmax}_j(\boldsymbol{c}_{j} \boldsymbol{v}^{\top}+\frac{\mu}{|t_v-t_j|}).
\end{equation}
Suppose $\tilde{\boldsymbol{v}}\in\mathbb{R}^{Kd}$ is the concatenation of $K$ $\boldsymbol{v}$s, $u$'s probability of interacting with $v$ is finally computed as
\begin{equation}\label{eq:y}
    \hat{y}_{uv} =\sigma(\tilde{\boldsymbol{u}}\tilde{\boldsymbol{v}}^{\top})= \sigma\bigg(\sum\limits_{j=1}^{K}\tilde{\boldsymbol{c}}_j\boldsymbol{v}^{\top}\bigg).
\end{equation}

The computation of $\hat{y}_{uv}$ implies that, the matching degree between $u$ and $v$ is determined by aggregating $v$'s matching to each factor of $u$'s preference.

 \vspace{-0.2cm}
\subsection{Model Optimization}
 \vspace{-0.1cm}
We adopt Bayesian personalized ranking (BPR) \cite{BPR} as the optimization algorithm to train our SR model. Specifically, suppose $\mathcal{N}_{u}$ is $u$'s interaction sequence fetched from the training set, the BPR loss of $u$'s training samples is computed by
\begin{equation}\label{eq:BPR}
 \vspace{-0.1cm}
\mathcal{L}_{BPR}=-\sum_{i\in \mathcal{N}_u}\sum_{j\notin \mathcal{N}_{u}} \ln\sigma(\hat{y}_{ui}-\hat{y}_{uj}).
\end{equation}

According to the primary principle of disentangled representation, each unit of the representation should be as independent as possible \cite{DGCF,DisenGCN}. It is required that all preference factors learned by the HGNN should be differentiable. According to this principle, node $i$'s assignment distribution vector $\boldsymbol{S}_{i}$ should be learned approximate to one-hot vector. In other words, node $i$ should be clustered into a factor node as rigidly as possible.
To this end, we add an entropy-based regularizer as follows:
\begin{equation}\label{eq:Ldis}
    \mathcal{L}_{Ent}= \frac{1}{N}\sum_{i=1}^{N}Entropy(\boldsymbol{S}_{i}) = -\frac{1}{N}\sum_{i=1}^{N}\sum_{j=1}^{K}\boldsymbol{S}_{ij}\log{\boldsymbol{S}_{ij}}.
\end{equation}

Finally, we have the following comprehensive loss function added with $L_{2}$ regularization $\|\Theta\|_{2}^{2}$, in which $\Theta=\{\boldsymbol{X}^{l},\boldsymbol{W}^{l}\} (1\leq l \leq L)$ and $M$ is the total number of users (sequences) in the training set,
\begin{equation}\label{eq:L}
 \vspace{-0.1cm}
\mathcal{L} = \sum_{u=1}^{M}\big(\mathcal{L}_{BPR} + \lambda_{1}\mathcal{L}_{Ent}\big) + \lambda_{2}\|\Theta\|_{2}^{2}.
 \vspace{-0.1cm}
\end{equation}

\vspace{-0.1cm}
\section{Evaluation}
\vspace{-0.1cm}
\nop{
In this section, we try to answer the following research questions (RQs) through our extensive experiments.

    \textbf{RQ1:} Can our SR model outperform the SOTA GNN-based SR models?
    
    \textbf{RQ2:} How do different components or settings of our model affect its SR performance?
    
    
   \textbf{RQ3:}  Can our proposed HGNN differentiate the different factors in a user's preference, resulting in better interpretability for our model's recommendation results?
}

\subsection{Datasets and Sample Collection}
\vspace{-0.1cm}
In our experiments, we evaluated all compared models on the following two multimedia recommendation datasets, of which the statistics are listed in Table \ref{tb:datasets}.

\textbf{Steam\footnote{http://cseweb.ucsd.edu/~jmcauley/datasets.html\#steam\_data}}:
This dataset is collected from Steam, which is a video game distribution platform, and contains reviews, timestamps and genres of massive games.

\textbf{MovieLens\footnote{https://grouplens.org/datasets/movielens/}}:
This is a benchmark movie recommendation dataset collected from MovieLens website. We adopted MovieLens-1M version in our experiments which contains ratings, timestamps and genres of various movies.

\begin{table}[!htb]
\vspace{-0.2cm}
    \caption{Dataset statistics.} \label{tb:datasets}
    \centering
     \resizebox{\columnwidth}{!}{
    \begin{tabular}
    {|ccccc|}
        \hline
        
        \hline
        Dataset  & Sequence \# & Item \# & Interaction \# & Genre \#  \\
        \hline
        Steam & 158,091 & 11,667 & 2,055,183 & 22  \\
        MovieLens & 74,079 & 3,390 & 963,027 &  18 \\
         \hline
         
         \hline
    \end{tabular}
    }
   \vspace{-0.1cm}
\end{table}

To collect samples, we sorted user interactions according to their timestamps to obtain an interaction sequence for each user. As in \cite{SRGNN}, we first filtered out the sequences with less than $N+1$ interactions, and then prepared each sample by a sliding-window. Concretely, we grouped the previous $N$ interactions as the input sequence, and took the last interaction as the ground-truth of next interaction. We used the earlier 90\% interactions as the training set, and the latter 10\% interactions as the test set. Then, the latter 10\% interactions in the training set were further used as the validation set. 
The sample processing file and our model's source-code are both provided in \url{https://github.com/xlx0010/HGNN}.

\vspace{-0.2cm}
\subsection{Compared Models}
\vspace{-0.1cm}
Although some of following baselines were designed for session-based recommendation, they were still compared since a truncated interaction sequence can be regarded as a session and there is limited work on GNN-based model specific to SR.

\noindent\textbf{FPMC} \cite{Rendle2010Factorizing}: It is a representative sequential model for SR built based on personalized Markov chain.

\noindent\textbf{GRU4REC+} \cite{hidasi2018recurrent}: It is an improved versions of GRU4REC \cite{GRUrec} using BPR as the loss.

\noindent\textbf{SASRec} \cite{SASRec}: It is a representative SR model with self-attentions.
        
\noindent\textbf{SR-GNN} \cite{SRGNN}: It employs gated graph neural network (GGNN) to capture the complex transition pattern in a session's interaction sequence.

\noindent\textbf{GC-SAN} \cite{GCSAN}: It adopts self-attention mechanism to learn sequence representations, where sequence graph structure and information propagation strategy are the same as SR-GNN.
    
\noindent\textbf{FGNN} \cite{FGNN}: GAT is employed in this SR model to capture transition patterns in sequence graphs. Specifically, FGNN assigns different weights to different neighbors, and uses weighted sum to update the central node instead of a complex gated mechanism. 

\noindent\textbf{RetaGNN} \cite{RetaGNN}: It trains a relational attentive GNN on a User-Item-Attribute tripartite graph and adopts
unidirectional self-attention to capture sequential characteristic in item sequence, but ignores timespan between items. For fair comparison, we removed its attribute nodes. 

Our SR model is denoted as \textbf{HGNN} and further compared with the following ablated variants.

\noindent\textbf{HGNN-GAT1}: 
The first GAT in HGNN is removed from this variant, resulting in $\boldsymbol{Z}$=$\boldsymbol{X}$.

\noindent\textbf{HGNN-GAT2}: 
 This variant was proposed by removing the second GAT in HGNN, resulting in the absence of assignment matrix $\boldsymbol{S}$. Namely, there is no node clustering ($\boldsymbol{C}$=$\boldsymbol{Z}$).
 
\noindent \textbf{HGNN-T}: 
In this variant, timespan is not included in the sequence graph. Accordingly, the timespan $\Delta t_{ij}$ is removed from relevant computations.
    
\noindent\textbf{HGNN-Ent}: 
It was proposed by removing the entropy-based regularizer $\mathcal{L}_{Ent}$ in our model's loss Eq. \ref{eq:L}. It was compared to verify the effect of differentiating preference factors on performance gains.


\vspace{-0.2cm}
\subsection{Experiment Settings}
\vspace{-0.1cm}
We used two typical SR metrics \emph{hit ratio (Hit)} and \emph{reciprocal rank (RR)} to evaluate all compared models. For fair comparison, all baselines' hyper-parameters were tuned to their optimal values against the datasets. Due to space limitation, we only display our model's results of $d=64$ and $N=12$ in the subsequent figures and tables. The consistent conclusions can still be drawn based on the results of other settings. Note that all models can not model user preferences well if $N$ is too small, and are time-consuming if $N$ is too large. In addition, all embeddings were initialized by a Gaussian distribution with a mean of 0 and a standard deviation of 0.1. 

In TSG construction, we set $T$=7 days (one week) for Steam and $T=30$ days (one month) for MovieLens, because interactions in Steam are denser in time than in MovieLens.
We set $\mu$=1 day for both datasets. In addition, we set $K$=5 and $\gamma$=0.8, according to our tuning results. The best setting of $K=5$ implies that the number of different preference factors of most users in the two datasets is within 5.

In addition, we used Adam \cite{Adam} optimizer with the learning rate of 0.001 and batch size of 1024, which were consistently used in all baselines. 
About the control parameter in Eq. \ref{eq:L}, we set $\lambda_1$ and $\lambda_2$ to $10^{-4}$ based on our tuning studies. All experiments were conducted on a workstation of dual GeForce GTX 1080 Ti with 32G memory and the environment of Ubuntu16.04 and torch1.7.1.

\begin{table*}[!htb]
    \caption{Performance comparison of all models on accuracy.} \label{tab:ex_results}
    \centering \small
  {
    \begin{tabular}
     {|c|cc|cc|cc|cc|}
       \hline
       
       \hline
        \multirow{2}{*} {\diagbox{\small{Model}}{\small{Dataset}}} &\multicolumn{4}{c|}{\textbf{Steam}}&\multicolumn{4}{c|}{\textbf{MovieLens}} \\ 
         \cline{2-9}
          &Hit@5 &RR@5 &\small{Hit@10} &\small{RR@10} &Hit@5 &RR@5 & \small{Hit@10} & \small{RR@10} \\ 
                \hline
                
                \hline
         FPMC        & 0.381 & 0.236 & 0.517 & 0.252 & 0.392 & 0.208 & 0.553 & 0.230\\
         GRU4REC+    & 0.391 & 0.253 & 0.530 & 0.272 & 0.462 & 0.274 & 0.629 & 0.303 \\
         SASRec     & {0.447} & {0.289} & {0.594} & {0.331} & {0.488} & {0.292} & {0.651} & {0.319} \\

         SR-GNN      & 0.561 & 0.314 & 0.705 & 0.338 & 0.459 & 0.241 & 0.628 & 0.280 \\
         GC-SAN      & 0.578 & 0.342 & 0.709 & 0.363 & 0.501 & 0.305 & 0.677 & 0.339 \\
         FGNN        & 0.602 & 0.385 & 0.725 & 0.392 & 0.538 & 0.324 & 0.701 & 0.354 \\
         RetaGNN     & \underline{0.622} & \underline{0.394} & \underline{0.734} & \underline{0.418} & \underline{0.561} & \underline{0.342} & \underline{0.747} & \underline{0.362} \\
             \%improvement   &3.70\%& 7.36\%& 3.01\%& 3.11\%& 4.46\%& 2.63\%& 2.95\%& 4.97\%\\ 
         \hline

         HGNN-GAT1   & \underline{0.615} & \underline{0.389} & \underline{0.731} & \underline{0.417} & \underline{0.543} & \underline{0.325} & \underline{0.746} & \underline{0.351} \\
         HGNN-GAT2   & 0.569 & 0.355 & 0.705 & 0.378 & 0.531 & 0.312 & 0.733 & 0.339\\
         HGNN-T      & 0.581 & 0.383 & 0.721 & 0.399 & 0.520 & 0.313 & 0.701 & 0.349\\
         HGNN-Ent    & 0.545 & 0.358 & 0.699 & 0.358 & 0.493 & 0.296 & 0.668 & 0.325\\
        \%improvement   &4.88\%& 8.74\%& 3.42\%& 3.36\%& 7.92\%& 8.01\%& 3.08\%& 8.26\%\\
        \hline

        \textbf{HGNN} & \textbf{0.645} & \textbf{0.423} & \textbf{0.756} & \textbf{0.431} & \textbf{0.586} & \textbf{0.351} & \textbf{0.769} & \textbf{0.380} \\
          \hline
          
          \hline
    \end{tabular}
    }
\vspace{-0.2cm}    
\end{table*}

\vspace{-0.2cm}
\subsection{Global Performance Comparisons}
\vspace{-0.1cm}
The comparison results of all models' performance are listed in Table \ref{tab:ex_results}, and HGNN's performance improvements over the strongest competitor (highlighted with underline) in the baseline and ablated version group are also displayed. The listed scores are the mean of five runnings for each model, verifying HGNN's remarkable superiority. Particularly, HGNNs superiority over the GNN-based SR baselines shows that, the hierarchical structure of sequence graphs is more beneficial than flat structure to capture the complex pattern in historical interactions. Although RetaGNN exhibits the SOTA baseline performance, it does not incorporate timespan and thus is still inferior to HGNN. 
In addition, GNN-based SR models exhibit superior performance compared with the sequential models (FPMC, GRU4REC+ and SASRec). 

\vspace{-0.2cm}
\subsection{Ablation Study}
\vspace{-0.1cm}
According to the ablation study results in Table \ref{tab:ex_results}, HGNN's advantage over HGNN-GAT1 shows the necessity of attentively aggregating neighbors' information to refine a node's embedding by GAT, which is crucial to better preference capture. HGNN's superiority over HGNN-GAT2 justifies the advantage of building the hierarchical structure for sequence graphs by node clustering. HGNN's superiority over HGNN-T verifies the significance of incorporating timespan into sequence graphs in terms of performance gains. In particular, HGNN's accuracy superiority in MovieLens is more apparent than Steam, because the average interaction timespan in MovieLens is bigger than Steam, making temporal information more significant. HGNN-Ent is also inferior to HGNN, because each factor is not distinguishable from others without the effect of entropy-based regularizer in HGNN-Ent. 

\vspace{-0.2cm}
\subsection{Case Study}\label{sec:case}
\begin{figure}[!htb]
\vspace{-0.1cm}
  \centering
  \includegraphics[width=3in]{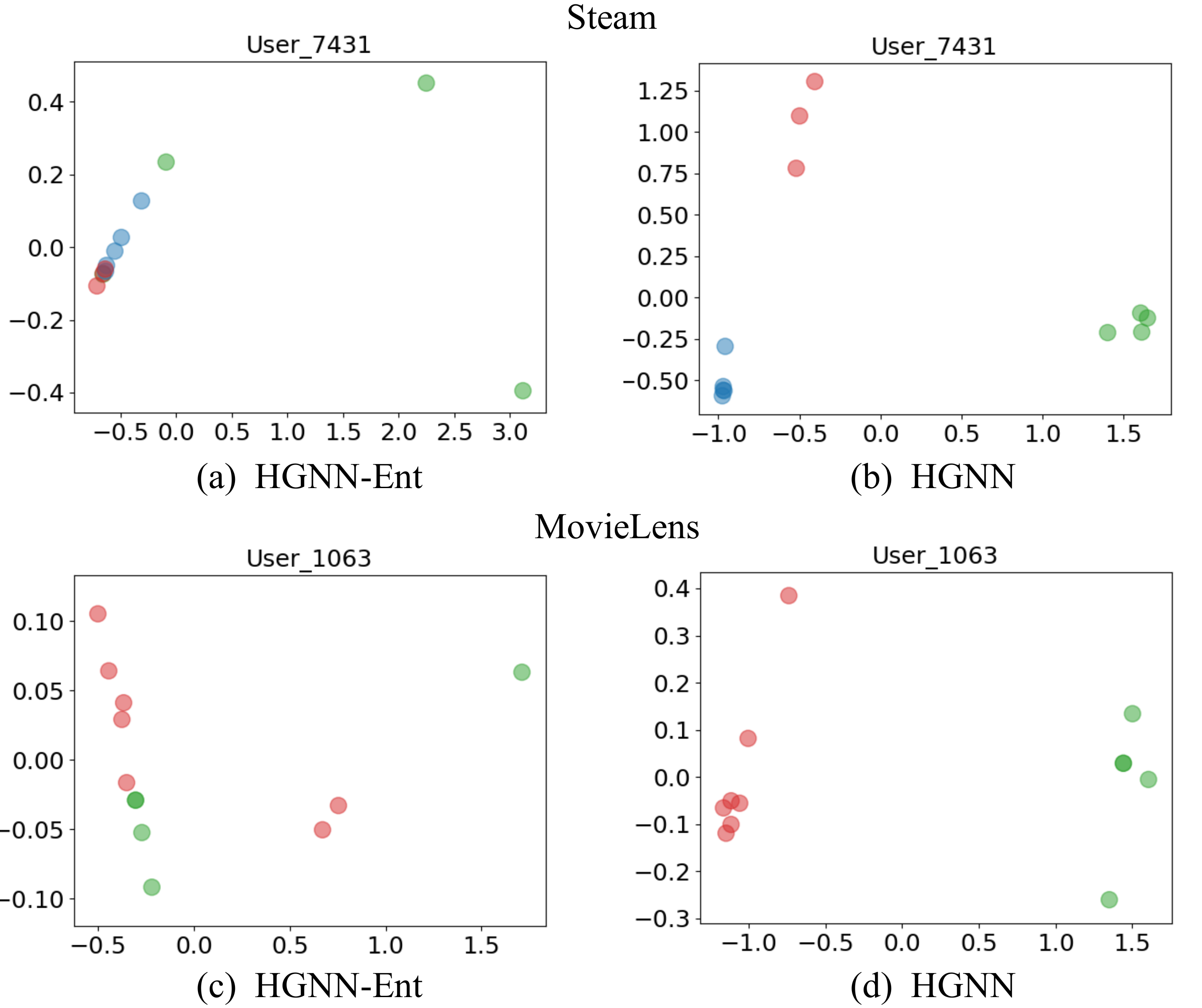}
  \vspace{-0.3cm}
  \caption{Node clustering results of (a/c) HGNN-Ent and (b/d) HGNN for two toy users (better viewed in color). 
 }\label{fig:cluster}
 \vspace{-0.4cm}
\end{figure}

\vspace{-0.1cm}
We further visualize some recommendation cases to demonstrate our model's advantages. To justify HGNN's advantage on differentiable clustering for capturing user factorial preferences, we rigidly categorized each node in TSG to one factor at first. Thus, the actual clusters may be less than 5 (we set $K$=5). Fig. \ref{fig:cluster} displays the positions of the 12 interacted items of one Stream user and one MovieLens user in 2D embedding space, which were learned by HGNN-Ent (Subfig. (a/c)) and HGNN (Subfig. (b/d)), respectively. In Fig. \ref{fig:cluster}, the nodes of each cluster are labeled with a certain color. We obviously find that HGNN clusters all nodes more differentially with the aid of entropy-based regularizer $\mathcal{L}_{Ent}$. Such results also ensure the rationality using the factorial preference representations as a user's disentangled representation.

\begin{figure}[!htb]
  \centering
  \includegraphics[width=3.5in]{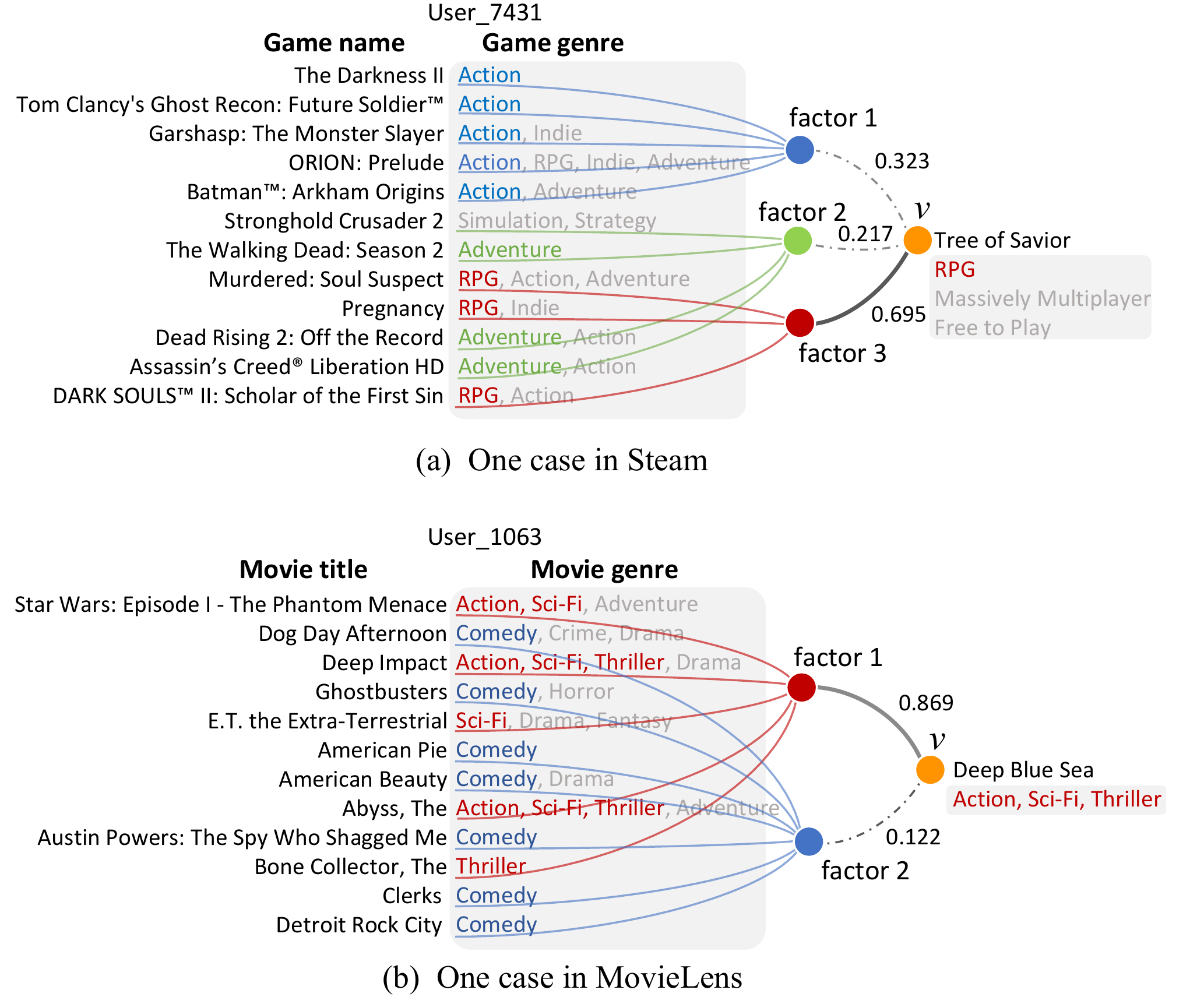}
  \vspace{-0.6cm}
  \caption{The shared genres (red font) between the predicted item $v$ (orange circle) and the historical interacted items belonging to main contributing factor (red circle) demonstrate HGNN's interpretability for recommendation results.}\label{fig:interp}
  \vspace{-0.4cm}
\end{figure}

To justify HGNN's interpretability for recommendation results, we compared the genres of historical interacted item and the predicted item. It is inspired by our empirical finding that item genres mostly match the factors of user preferences in these two multimedia datasets. Fig. \ref{fig:interp} displays the same two cases in Fig. \ref{fig:cluster}, where the 12 interacted items are listed chronologically in the left column. We highlighted the item genres that represent user preference factors mostly. The right orange circle represents the next interacted item $v$ predicted by our model. The edge scores were computed in Eq. \ref{eq:y}, indicating the significance (correlation) of each preference factor to $v$. It shows that $v$ has the same genres as those of historical interacted items belonging to the same factor (factor 3 for User\_7431 and factor 1 for User\_1063). Such correlations provide a persuasive reason for predicting $v$, demonstrating our model's interpretability.

\vspace{-0.1cm}
\section{Related Work}
\vspace{-0.2cm}
Many traditional SR models adopt Markov chains to capture sequential patterns between consecutive items in a sequence \cite{Huang2009Markov}. Afterwards, FPMC \cite{Rendle2010Factorizing} combines Matrix Factorization (MF) and Markov Chain to model sequential behaviors, of which a major problem is the static representations for user intentions. Recently, some Deep Neural Networks (DNNs) such as LSTM and GRU have been employed in SR models to capture user preferences through encoding historical interactions into a hidden state. 
Among them, GRU4REC \cite{GRUrec} is the pioneer work, which encodes items into one-hot embeddings and feeds them into GRUs to achieve recommendation. Then, GRU4REC+ \cite{hidasi2018recurrent} was proposed as an advanced version of GRU4REC. 

Inspired by the power of GNNs \cite{GCN,GAT} on graph modeling, some efforts employed GNNs to guide the learning of user/item representations for CF-based performance gains. For example, GC-MC \cite{GCMC} applies graph convolutional network (GCN) \cite{GCN} on user-item graph. For GNN-based SR models, Wu et al. proposed SR-GNN \cite{SRGNN} which uses gated graph neural network (GGNN) \cite{GGNN} to model the sequence graph, and thus the complex transition pattern rather than the sequential transition pattern is captured. Besides SR-GNN, GC-SAN \cite{GCSAN} and MKM-SR \cite{MKM} also employ GGNN to capture the complex transition pattern in sequence graphs to achieve enhanced SR. Similar to our model, FGNN \cite{FGNN} utilizes GAT to capture the item transitions in sequence graph. SURGE \cite{SURGE} constructs an interest graph for a given user also through GNN-based clustering algorithm. We believe it is inferior to our model since it excludes temporal information during its graph construction, although we did not compare it in our experiments due to the lack of source code.
\vspace{-0.1cm}
\section{Conclusion}
\vspace{-0.1cm}
We propose a novel SR model in which the user's sequence graph is constructed into a TSG with the timespans between interacted items. Our proposed HGNN is then applied on this graph to capture the complex pattern and differentiable factors in user preferences, resulting in enhanced SR performance.  
Our extensive experiments not only justify our model's performance advantages over SOTA models but also verify the necessity of our model's components. 
\vspace{-0.1cm}

\vspace{-0.2cm}
\bibliographystyle{IEEEbib}
\small
\bibliography{refer}

\end{document}